\documentclass[fleqn,usenatbib]{mnras}

\usepackage[T1]{fontenc}

\DeclareRobustCommand{\VAN}[3]{#2}
\let\VANthebibliography\thebibliography
\def\thebibliography{\DeclareRobustCommand{\VAN}[3]{##3}\VANthebibliography}

\usepackage{graphicx}	
\usepackage{amsmath}	
\usepackage{amssymb}	

\newcommand{\MSOL}{\mbox{$\:M_{\sun}$}}

\newcommand{\EXPU}[3]{\mbox{\rm $#1 \times 10^{#2} \rm\:#3$}}  

\title{An Improved Model for the Spectra of Disks of Nova-like Variables}

\author[Hubeny and Long]{
Ivan Hubeny$^{1}$\thanks{E-mail: ihubeny.astr@gmail.com}
and Knox S. Long$^{2,3}$
\\
$^{1}$The University of Arizona, Steward Observatory, 933 North Cherry Avenue, Tucson, AZ 85719, USA\\
$^{2}$Space Telescope Science Institute,
3700 San Martin Drive,
Baltimore MD 21218, USA; long@stsci.edu\\
$^{3}$Eureka Scientific, Inc.
2452 Delmer Street, Suite 100,
Oakland, CA 94602-3017, USA
}

\date{Accepted XXX. Received YYY; in original form ZZZ}

\pubyear{2021}

\begin{document}
\label{firstpage}
\pagerange{\pageref{firstpage}--\pageref{lastpage}}
\maketitle

\begin{abstract}
The spectra arising from the  disks of nova-like variables show many of the features seen in stellar atmospheres. They are typically modelled either from an appropriated weighted set of stellar atmospheres or a disk atmosphere with energy is dissipated near the disk plane, with the effective temperature distribution expected from a steady state accretion disk.  However these models generally over-predict the depth of the Balmer jump and the slope of the spectrum in the ultraviolet.  The problem is likely due to energy dissipation in the disk atmosphere,  which produces a flatter vertical temperature profile than is observed in stars.  Here, we provide validation for this hypothesis in the form of spectra generated using the stellar atmosphere code {\sc tlusty} using  a parametric prescription for energy dissipation as a function of depth and closely match the spectrum of the nova-like IX Vel over the wavelength range 1150-6000 \AA.
\end{abstract}




\section{Introduction}

Cataclysmic variables are binary star systems with typical periods of about 80 minutes \citep{gansicke09} to greater than one day \citep{ritter03},  consisting of a white dwarf (WD) that accretes matter via Roche lobe overflow from a relatively normal secondary star \cite[see, e. g.][for a review]{warner95}.  Although some such systems have WDs with strong magnetic fields, which funnel matter from the secondary on the small foot points on the WD,  accretion in most systems is mediated by an accretion disk. A large fraction of the disk-dominated systems, the so-called dwarf novae, undergo quasi-periodic eruptions where the disk transitions from a relatively cold, low mass transfer state to a hot, high mass-transfer rate state.  A smaller fraction, the nova-like variables, remain in the hot, high mass transfer rate all, or most of the time.  Many astrophysical systems contain accretion disks, of course, from young stellar objects to active galactic nuclei, and these accretion disks manifest themselves in various ways in the spectra observed from such systems.  Dwarf novae and nova-like variables are the closest (and most ubiquitous) examples of accretion disk systems.  They radiate mostly in the UV and visible wavelength range.  The properties of the primary and secondary are well understood.  As a result, CVs provide an ideal laboratory for understanding accretion disk physics.   

As first studied with IUE, most non-eclipsing nova-like variables and dwarf novae in outburst have ``blue'' spectra in the UV \cite[see, e.g.][]{krautter81}, and have usually been interpreted within the context of models  developed for steady state disks  by  \cite{shakura73}
and \cite{lynden74} by summing an appropriate weighted set of blackbodies or stellar atmospheres \cite[][and references therein]{wade84}.   Both approaches to fitting the observed FUV spectra  yielded reasonable agreement with the  overall spectral slope and in the case of the summed stellar atmospheres approach reproduce some of the features in the spectra even when higher signal to noise spectra were obtained with instruments like the Hopkins Ultraviolet Telescope  \citep{long91,long94} or the various spectrographs on the Far Ultraviolet Spectroscopic Explorer and Hubble Space Telescope \citep{linnell_mvlyrae,linnell_ixvel,linnell_v3885sgr,linnell10,godon20}.  

Anomalies were apparent however when broader spectra ranges were considered.  In particular the spectra of nova-like variables and dwarf novae in outburst show Balmer jumps.  And while synthetic spectra spectra constructed from summed blackbodies have, by construction, no Balmer jumps,  synthetic spectra created with summed stellar atmospheres have Balmer jumps that are more pronounced than observed. On the other hand, BB spectra reproduce the spectral slope over the FUV/NUV wavelength range more accurately than those produced with summed stellar atmospheres \citep{wade88}.  

 A variety of suggestions have been made to address the problem of the spectral slope.  Most involve modifying the run of  temperature with radius in the disk \citep{linnell10} or ``removing'' a portion of the inner disk \citep{linnell_mvlyrae,linnell_ixvel}. Some, such as \cite{kromer07} have suggested that irradiation could modify the vertical temperature profile of the disk. Others suggested that emission from the winds known to emanate from the disks of high state CVs would work \citep{knigge97,matthews15}  or that winds carried away enough mass to alter the temperature profile \citep{knigge99}.  A recent example of the later is the work by \cite{nixon19}, who suggest that in nova-like variables large scale magnetic fields modify the disk temperature.

However, it also was recognized fairly early on that differences between the structure of the disk and that of a stellar atmosphere might account for these problems, particularly because viscous energy is likely to be released  in the region of the photosphere of a disk, unlike stars where the energy release occurs deep beneath the atmosphere.  \cite{shaviv91}, in particular, argued, based on their own purpose-built radiative transfer code, that this was the main reason that the Balmer jump was shallower in in nova-like variables than expected based on models constructed from stellar atmospheres.\footnote{Unfortunately, there was very little followup to this work. \cite{idan10} describe an improve version of the code used by \cite{shaviv91}, but it is not clear that the claim that the Balmer jump is reduced in the improved models.  The synthetic spectra produced there have obvious Balmer jumps (see their Fig.\ 9, but tno comparison to observed spectra of nova-likes or other high state cataclysmic variables is made and there is no explicit discussion of the Balmer region of the spectrum. They say their model spectra are similar to those of \cite{wadehub98} which were created with {\sc tlusty} assuming all of the viscous energy was deposited deep within the atmosphere, and have Balmer jumps that are stronger than observed.}

Despite these early indications that viscous dissipation was the likely explanation of the Balmer jump problem and the existence of better flux calibrated and higher S/N spectra of CVs, there have been few, if any, 21st century attempts to model the spectra of disk-dominated CVs in the high mass transfer state using modern codes capable of handling dissipation through out the disk atmosphere.  Part of the reason for this lack of progress is that we still do not have a physical model of how dissipation occurs in the upper layers of the disk.  In its absence, however, it should be possible to use a parameterized dissipation formula that in conjunction with a ``stellar atmospheres'' code  allows one to create simulated spectra of CV accretion disks for comparison with observations.  

Here we provide just such a procedure. For this purpose we use the latest version of code  {\sc tlusty} \cite[][]{hubeny88,hubeny95} which is capable of modelling the atmospheres of both stars and accretion disks.\footnote{The code was originally existed in two flavors  {\sc tlusty} for modelling stellar atmospheres and one called {\sc tlusdisk}, but these were later merged into a single coce {\sc tlusty} }  Various versions of {\sc tlusty} have been used in the majority of attempts to model CV spectra,  but it was not possible at the time of those attempts to explore the types of dissipation laws that are now possible.   Although the formalism contained in  {\sc tlusty} then allowed for viscous dissipation though out the disk atmosphere, the interaction of radiation and matter,
albeit treated self-consistently by solving the radiative transfer equation together with other structural
equation, was still relatively simplistic. As a result, a wide range of dissipation laws could not be explored. Consequently, all of the attempts to model spectra to date have either been carried out using either a stellar atmospheres approach \citep{long91,long94}, or a disk approach with all of the viscous dissipation at depths well below the photosphere \citep{linnell_ixvel,linnell_v3885sgr,linnell10,godon20}.

The remainder of this report is organized as follows:  In \S\,\ref{model}, we summarize the model assumptions we adopt, and the basic structural equations to compute the vertical structure of an accretion disk. 
In \S\,\ref{tlusty}, we describe the improvements that have been made to {\sc tlusty} that enable the current approach.  In \S\,\ref{sensitivity}, we present a sensitivity study
with the aim to show how the basic parameters used to described local viscosity influence the resulting 
temperature structure and emergent spectrum. In \S\,\ref{ixvel}, we present an application of our models
to the popular and well-studied CV system IX Vel.  
In \S\,\ref{concl}, we summarize our conclusions.
Finally, Appendices \ref{col_mass}  and \ref{rybicki}  present some numerical details of model construction.

\section{Model assumptions and basic equations}
\label{model}

We adopt here the standard assumptions used to construct models of vertical structure of accretion disks
in cataclysmic variable systems. The disk is divided into a set of concentric rings, each behaving
as an independent 1-D plane parallel radiating slab. Each ring is assumed to be in hydrostatic equilibrium in 
the vertical direction. Energy balance is considered as a balance between the net radiation
loss, calculated exactly, without invoking either optically thin or optically thick (diffusion) approximations. 
Mechanical energy is generated through viscous dissipation. The dissipated energy is proportional to viscosity, 
which is given through the empirical viscosity parameter $\alpha$ \citep{shakura73}.

The basic structural equations adopted in our modeling procedure were described in detail by
\cite{hubeny90}, \cite{hubeny98}, and \cite{hubeny17}. 
Here we provide a brief summary.
We consider only non-relativistic disks, although our approach can be adapted for relativistic disks as well \citep{hubeny98}.

The atmosphere at each disk radius $R$ (specified at the disk
central plane) is in vertical hydrostatic equilibrium, with a depth-dependent gravity
($g$) that arises from the vertical component of the central star's
gravitational force on the disk material. Neglecting the self-gravity of
the disk and assuming that $R$ is much larger than the distance from the
central plane, $z$,  the pressure $P$ satisfies
\begin{equation}
\label{hedisk}
\frac{dP}{dz} = -g(z)\rho, \quad {\rm or} \quad \frac{dP}{dm} =  g(z)\, ,
\end{equation}
where the depth-dependent vertical gravity acceleration is given by
\begin{equation}
\label{gdef}
 g(z) = \frac{GM} {R^3} \, z \, .
\end{equation}
where $G$ is the gravitational constant, $M$ the mass of the central star, and $m$ the column mass, given by $dm = -\rho dz$.

The energy balance equation is written as
\begin{equation}
\label{ebal}
\frac{\partial F_z}{\partial z} = \frac{3}{2}\left(\frac{GM}{R^3}\right)^{1/2}\, t_{\phi r},
\end{equation}
where $F_z$ is the $z$-component of the energy flux and $t_{\phi r}$
is the sheer stress, also called the viscous stress.
Under the assumption of $t_{\phi r}=0$ at the innermost orbit, one obtains from 
momentum conservation,
\begin{equation}
\label{mbal}
\int_{-h}^{h} t_{\phi r} dz = \frac{\dot M}{2\pi}\left(\frac{GM}{R^3}\right)^{1/2}
[1-\left({R_\ast/R}\right)^{1/2}],
\end{equation}
where $h$ is the vertical height of the given annulus, 
and $\dot M$ is the mass accretion rate.
The source of viscous stress is described by
\begin{equation}
\label{vbal}
t_{\phi r}  = \frac{3}{2}\eta\left(\frac{GM}{R^3}\right)^{1/2},
\end{equation}
where $\eta$ is the coefficient of sheer viscosity, which is expressed through
the coefficient of kinematic viscosity $w$ as $\eta\equiv\rho w$.

To write down practical expressions for the energy balance and for the
total column mass, one has to introduce a suitable parametrization of viscosity.
The vertically averaged kinematic viscosity is given by
\begin{equation}
\label{avervisc}
\bar{w} = \frac{\int_0^h w \rho\, dz}{\int_0^h \rho dz} = \frac{1}{m_0} \int_0^h \eta \,dz =
\frac{1}{m_0} \int_0^{m_0}\!\! w\, dm.
\end{equation}
Integrating Eq. (\ref{vbal}) from 0 to $h$, and using Eq. (\ref{mbal}) together
with Eq. (\ref{avervisc}), one can express the total column mass at the central plane
through the averaged viscosity as
\begin{equation}
\label{m0}
m_0 = \frac{1}{\bar w} \frac{\dot M}{6\pi} [1-\left({R_\ast/R}\right)^{1/2}]
\end{equation}

To express the viscosity, we use commonly used prescription is based on the so-called
$\alpha$-parametrization \citep{shakura73}. There are several variants
of this parametrization; we use here a version in which the vertically
averaged sheer viscosity is taken proportional to the vertically averaged (total) pressure,
\begin{equation}
\bar{t_{\phi r}} \equiv \frac{1}{h} \int_0^h t_{\phi r} dz = \alpha \bar P,
\end{equation}
in which case
\begin{equation}
\label{inttfir}
\int_0^h t_{\phi r} dz = h \alpha \bar P = m_0 \alpha (\bar P/\bar\rho),
\end{equation}
where the averaged density is given by $\bar\rho = m_0/h$. The vertically
averaged kinematic viscosity is given, substituting Eqs. (\ref{vbal}) integrated over
$z$ into Eq. (\ref{inttfir}), by
\begin{equation}
\bar w = \alpha\, \frac{2}{3} \left(\frac{R^3}{GM}\right)^{\!1/2} 
\left(\frac{\bar P}{\bar\rho}\right).
\end{equation}
A numerical disadvantage of the $\alpha$-prescription is that the vertically averaged kinematic
viscosity, and the total column mass, are not known a priori since they depend on
$(\bar P/\bar\rho)$, which can only be accurately computed when the model is
constructed. To cope with this problem, we have developed an iterative procedure  to determine the column mass. This procedure is described in Appendix \ref{col_mass}.

Integrating Eq.\,(\ref{ebal}) over $z$, and using Eq.\,(\ref{mbal}), one obtains for the total energy flux at the surface, which is expressed, in analogy to stellar atmospheres, through the effective temperature,
\begin{equation}
\label{teff_def}
F_z(h) \equiv \sigma_R T_{\rm eff}^4\  =\ 
{3\over 8\pi} {G M \dot M \over R^3}\,   [1-\left({R_\ast/R}\right)^{1/2}],
\end{equation}
The integral form of the energy balance equation follows directly from
Eqs. (\ref{ebal}) and  (\ref{vbal}):
\begin{equation}
\label{ebal_int}
4\pi \int_0^\infty (\eta_\nu- \kappa_\nu J_\nu)\, d\nu = 
\frac{9}{4} \frac{GM}{R^3} \rho w,
\end{equation}
where the left-hand side, analogous to the case of stellar atmosphere, expresses
the net energy radiated away per unit volume, while the right-hand side 
expresses the total energy generated by viscous dissipation in the same unit
volume. Here, $\eta_\nu$ and $\kappa_\nu$ are the emission and absorption coefficients, respectively, at frequency $\nu$, and $J_\nu$ is the mean intensity of radiation. The differential form, in analogy with the atmospheric case, is written as:
\begin{equation}
\label{ebal_dif}
4\pi\int_0^\infty \frac{d(f_\nu J_\nu)}{d\tau_\nu}\, d\nu = \sigma_R T_{\rm eff}^4
[1-\theta(m)],
\end{equation}
where $f_\nu$ is the Eddington factor 
[see Eq. (\ref{vef}) below], and the function $\theta$ is defined by
\begin{equation}
\label{theta}
\theta(m) \equiv \frac{1}{\bar m_0} \int_0^m w(m^\prime)\,dm^\prime,
\end{equation}
which is a monotonically increasing function of $m$ with $\theta(0)=0$ and
$\theta(m_0)=1$, for any dependence of the local viscosity on depth.

As in the case of stellar atmospheres, we use here a linear combination of
Eqs. (\ref{ebal_int}) and (\ref{ebal_dif}), which exhibits much  more stable numerical behavior.
 If convection is present, the convective
flux $F_{\rm conv}$  is added, as is the case for stellar atmospheres.
The complete energy balance equation then reads:
\begin{eqnarray}
\label{re_con_disk}
a\bigg[\int_0^\infty\!\!\left(\kappa_\nu J_\nu - \eta_\nu\right)d\nu +E_{\rm diss}
+ \frac{\rho}{4\pi}\frac{dF_{\rm conv}}{dm}  \bigg] + \nonumber \\
b \bigg[\int_0^\infty\! \frac{d(f_\nu J_\nu)}{ d\tau_\nu}\,d\nu -
\frac{\sigma_R }{ 4\pi}\,T_{\rm eff}^4[1-\theta(m)]
+ \frac{F_{\rm conv}}{ 4\pi}\bigg] = 0,
\end{eqnarray}
where
\begin{equation}
\label{ediss}
E_{\rm diss} = \frac{9}{16\pi} \frac{GM}{R^3}  \rho w,
\end{equation}
is the total energy generated by viscous dissipation per unit volume. 
Empirical parameters $a$ and $b$ are chosen  such as $b=0$ at
low optical depths, because the radiation intensity changes very little, so that a numerical
evaluation of the derivative in Eq. (\ref{ebal_dif}) is inaccurate and unstable. Similarly, the parameter $a=0$ 
at the deep layers.

Finally, the radiation field and its interaction with matter is described by the radiative transfer equation, conveniently
written in  the second-order form,
\begin{equation}
\label{rte}
\frac{d^2 (f_\nu J_\nu)}{d\tau_\nu^2} = J_\nu - S_\nu,
\end{equation}
where  $\tau_\nu$ the monochromatic optical depth, and $S_\nu$ the source function, defined by
\begin{equation}
\label{sf}
S_\nu = \frac{\eta_\nu + \sigma_\nu J_\nu}{\kappa_\nu +\sigma_\nu}
\end{equation}
where $\sigma_\nu$ is the coefficient of scattering. Here we assume that the scattering process is isotropic.
The Eddington factor is defined by
\begin{equation}
\label{vef}
f_\nu\equiv K_\nu/J_\nu=\int_{-1}^1 I_\nu(\mu)\mu^2\,d\mu \bigg/ 
\int_{-1}^1 I_\nu(\mu)\,d\mu,
\end{equation}
where $I_\nu$ is the specific intensity of radiation,
$\mu$ is the cosine of the angle between the direction of
propagation of the radiation and the normal to the surface.
It is obtained by a formal solution of the angle-dependent radiative 
transfer equation -- for details refer, e.g., to \cite{hubmih14}.
The optical depth is defined by
\begin{equation}
d\tau_\nu \equiv -\chi_\nu dz = (\chi_\nu/\rho)\, dm,
\end{equation}
where $\chi_\nu = \kappa_\nu + \sigma_\nu$ is the total absorption (also called extinction) coefficient.

The upper boundary condition is written as
\begin{equation}
\label{rte_ubc}
\left[\frac{\partial(f_\nu J_\nu)}{\partial\tau_\nu}\right]_0 = g_\nu J_\nu(0)
- H_\nu^{\rm{ext}},
\end{equation}
where $H_\nu^{\rm{ext}}$ is the external irradiation flux, which we take here as  $H_\nu^{\rm{ext}}=0$, and
$g_\nu$ is the surface Eddington factor defined by
\begin{equation}
\label{vefg}
g_\nu\equiv \frac{1}{2}\int_{-1}^1 I_\nu(\mu,0)\mu\,d\mu \big/ J_\nu(0),
\end{equation}
and the lower boundary condition that represents a symmetry condition
at the midplane,
\begin{equation}
\label{rte_lbcd}
H_\nu = 0, \quad {\rm or} \quad \frac{d(f_\nu J_\nu)}{d\tau_\nu} = 0.
\end{equation}

The set of basic structural equations 
(\ref{hedisk}), (\ref{re_con_disk}), and (\ref{rte})--(\ref{rte_lbcd}) is complemented by the
charge conservation equation and  the relation between $m$ and $z$. For more details, refer to \cite{hubeny17}.


\subsection{Parameterization of the local viscosity} 

We adopt the  parameterization of the local viscosity using the formalism of \cite{hubeny98}.
The (depth-dependent) viscosity $w$ is generally allowed to vary as a step-wise 
power law of the mass column density, viz. 
\begin{eqnarray}
\label{visc1}
w(m) &=& w_0 \left( {m/m_0} \right)^{\zeta_0}\, , \quad  m>m_{\rm d}, \nonumber \\
w(m) &=& w_1 \left( {m/m_0} \right)^{\zeta_1}\, , \quad  m<m_{\rm d}  ,
\end{eqnarray}
where $m_{\rm d}$ is the division point.
In other words, we allow for a different power-law exponent for inner and
outer layers, and also for a different portion of the total energy dissipated in these
layers. This represents a generalization
of \cite{kriz86}, based on a single power-law representation.

We stress that this parameterization is an empirical one. The most natural way
of treating local viscosity would be to keep the coefficient of kinematic viscosity
constant with depth, i.e., $\zeta_0=\zeta_1=0$. 

Here we use for simplicity a step-wise constant dissipation, $\zeta_0=\zeta_1=0$,
so that $w(m)=w_0$ for $m>m_{\rm d}$, and is equal to $w_1$ elsewhere. These values
are not taken as free parameters. What is parameterized as a fraction, $f$, of the total energy
that is dissipated in  deep layers, $m>m_{\rm d}$. The coefficients $w_0$ and $w_1$
are then given by (for $\zeta_0=\zeta_1=0$; for general expressions refer to \cite{hubeny98}
\begin{eqnarray}
\label{visc2}
w(m) &=& w_0 = f\bar{w}/(1-m_d/m_0), \quad m>m_{\rm d} \nonumber \\
w(m) &=& w_1 = (1-f)\bar{w}/(m_d/m_0), \quad  m<m_{\rm d}.
\end{eqnarray}
Therefore, the values of $m_d$ and $f$ are the adopted free parameters (besides $\alpha$).


\section{Computational procedure}
\label{tlusty}

As mentioned above, we use  {\sc tlusty} to solve the coupled set of basic structural equation described above.
Most of the attempts to model the disks of CVs with {\sc tlusty} occurred ten or more years ago. Several significant improvements to the code have taken place since then.  

The first, relatively minor but still important,
improvement is a possibility of using $\alpha$, instead of the Reynold number, to parameterize the vertically-averaged
viscosity. The procedure is described in Appendix \ref{col_mass}.

The second, and in fact crucial, improvement
is an application of the so-called Rybicki scheme, originally developed by \cite{rybicki71} in the context of two-level
atom, and later adapted for use in computing model atmospheres - see \cite{hubmih14}, and \cite{hubmn17}
In the context of {\sc tlusty}, the method is described in detail by \cite{hubeny17}.
Some details of the implementation of the Rybicki scheme in the case of accretion disks is described in Appendix \ref{rybicki}.

The essential advantage of the Rybicki scheme is that one solves simultaneously the radiative transfer equation together
with the energy balance using  a linearization method. 
Thanks to a judicious reorganization of the global solution matrix, the method retains favorable numerical
properties of the complete linearization, namely a fast convergence and stability. At the same time it allows 
for a large number of discretized frequency points to describe the opacity and the radiation field.  Unlike the original complete linearization approach, in which computer time scales roughly  as a cube of the number of frequency points, in the Rybicki scheme computer time scales linearly with the number of frequencies.

The third main improvement is an application of opacity tables. Rather than computing the opacity on the fly as done previously, the current version of {\sc tlusty} interpolates from  a pre-calculated opacity table, generated by the accompanying
program {\sc synspec}.  A more detailed description of the code, and the issues associated with generating and using opacity tables, we refer the reader to  \cite{hubeny21}. The big advantage of
this approach is that one can include all necessary opacity sources at no additional costs for computing
model atmospheres or disks once the table is being generated.

Since the opacity table, in order to be sufficiently robust, has to contain opacities at many frequency points, the use
of the Rybicki scheme is more or less mandatory when working with opacity tables. One can in principle use
the hybrid Complete Linearization/Accelerated Lambda Iteration (CL/ALI) method \citep{hubeny95}. However, while
this method works well for NLTE model atmospheres of hot stars, it converges slowly, or, worse,  fails to converge
for conditions not in pure radiative equilibrium, including cool stars with convection or accretion disks.
This is because the radiation intensities at most frequency points are not explicitly linearized, so their feedback
to the energy equation is only taken one iteration step later.

There is an additional  significant benefit of using the opacity tables in the case of accretion disks. 
The previous {\sc tlusty} models considered only
a limited set of opacities, typically just hydrogen and helium bound-free and free-free transitions.
The opacities due to other species were added only in the solution of the radiative transfer equation after the structure
had been calculated. This led to numerical problems in the upper layers of a disk. This is essentially
because when going to upper layers, energy is still being generated there, bur the opacity is increasingly dominated by electron scattering, which is not an efficient process for radiating away the energy released.
That basically leads to a ``corona", as suggested by \cite{shaviv86}. 

To circumvent this problem \cite{kriz86} (and by default others who used {\sc tlusty} subsequently) resorted to  an ad-hoc decrease of dissipation toward the surface. However, with many opacity sources, that is no longer necessary.  Instead,
the numerical instability is avoided in a physically realistic way. One can
even consider models, like some  we describe here,  where dissipation increases {\em outward}.  This would not   be possible without considering realistic opacities,  produced by all continua and lines of all important atomic and molecular species and incorporated into an opacity table.


\section{Results}
\label{sensitivity}

\subsection{Basic parameters}

To examine the sensitivity of models and emergent spectra, with the emphasis on the Balmer edge region, on the adopted values of $m_d$ and $f$, we take as a test example an annulus of a disk around a white dwarf with mass $M_\ast = 0.8 M_\odot$, and the corresponding radius $R_\ast = 7.252\times 10^8$ cm, at a distance $R= 15\, R_\ast$, and with mass accretion rate $\dot M = 10^{-8} M_\odot$/yr.  The viscosity parameter $\alpha$ is taken at its typical value $\alpha=0.1$.
With these parameters, the effective temperature
of the annulus is $T_{\rm eff}=17,407$ K, and effective  $\log g$ (a value of a depth-dependent gravity acceleration
at $\tau_{\rm Ross} \approx 2/3$) being approximately $\log g_{\rm ref} \approx 5$, so that NLTE effects are
to a reasonable approximation negligible. We shall therefor consider only LTE models.

We use a table that includes
the line and continuum opacity of the following species: H I, He I-II, C I-IV, N~I-V, O I-VI, Ne I-V, Na I-II,
Mg I-II, Al I-III, Si I-IV, S II-V, and Fe I-VII. For temperatures below 8,000 K we include molecular line opacity
for a number of diatomic molecules \citep{hubeny21}.
The table was constructed for 21 temperatures, logarithmically spaced  between 2,000 and 200,000 K,
and 17 densities  between $10^{-12}$ and $10^{-3}$ g cm${}^{-3}$, for 30,000 wavelength points
between 200 and 100,000 \AA.  We have also constructed a table with the same parameters.
but for 100,000 wavelength points, but the results using this extended table were hardly distinguishable from those
obtained with the current table, while the model
construction took about three time more computer time, so we decided to use the 30,000 point table 
for all subsequent simulation.

\subsection{Sensitivity of models on $f$ and $m_d$}

%
\begin{table}
\caption{Input parameters for models with constant $\alpha=0.1$}
\begin{center}
\begin{tabular}[t]{|l|c|c|}
\hline
 Name & fraction $f$ &  division point $m_d/m_0$\rule{0in}{3ex}\\[1ex]
\hline
a0    &    0.99            &  $10^{-2}$  \\
a1    &    0.3              &  $10^{-2}$  \\
a2    &    0.1              &  $10^{-2}$  \\
a3    &    $10^{-2}$    &  $10^{-2}$  \\
a4    &    $10^{-3}$    &  $10^{-2}$  \\
\hline
b1    &    $10^{-2}$    &  $10^{-2}$  \\
b2    &    $10^{-2}$    &  $3\times10^{-3}$  \\
b3    &    $10^{-2}$    &  $10^{-3}$  \\
b4    &    $10^{-2}$    &  $3\times10^{-4}$  \\
b5    &    $10^{-2}$    &  $10^{-4}$  \\
\hline
\end{tabular}
\end{center} 
\end{table}  

For the above parameters of the ring, we explore a range of values of $m_d$ and $f$, listed in Table 1.
 
To provide a more insight into the results, and to explain a choice of adopted viscosity parameters, we
display the temperature not as a function of the Lagrangian column mass $m$, which is
a natural depth coordinate of the problem, but rather as a function of the Rosseland optical depth. We stress that
a dependence of the Rosseland optical depth on column mass is different for different models. In order to appreciate the role the division mass $m_d$ may play in determining the emergent radiation, we mark the positions of the individual division points by black dots. This helps to explain how the adopted value of $m_d$ may influence the emergent radiation. 

\begin{figure}
\begin{center}
\includegraphics[width=3.5in]{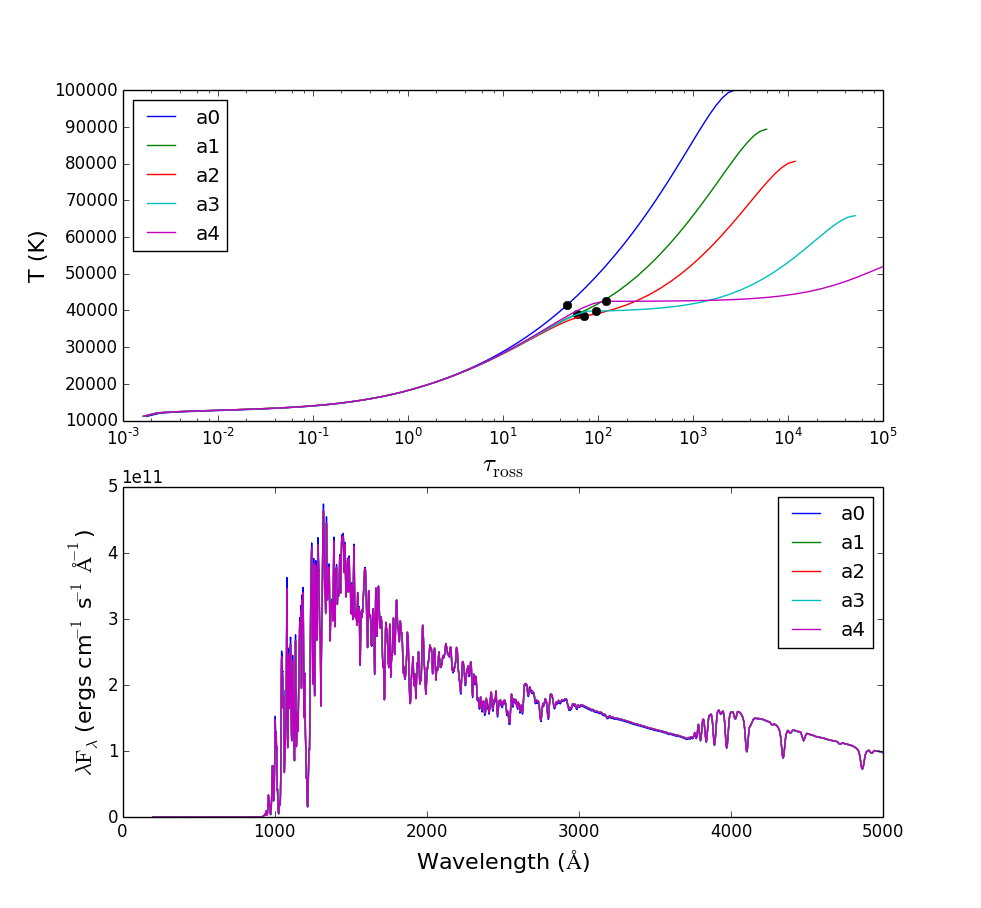}
\caption{Models a0 -- a4: Upper panel: temperature as a function of Rosseland optical depth. Black dots mark the position of the division point $m_d$. Lower panel: emergent flux at the disk surface. The flux is essentially the same for all models.
\label{fig:fig1}}
\end{center}
\vspace{-1em}
\end{figure}

\begin{figure}
\begin{center}
\includegraphics[width=3.5in]{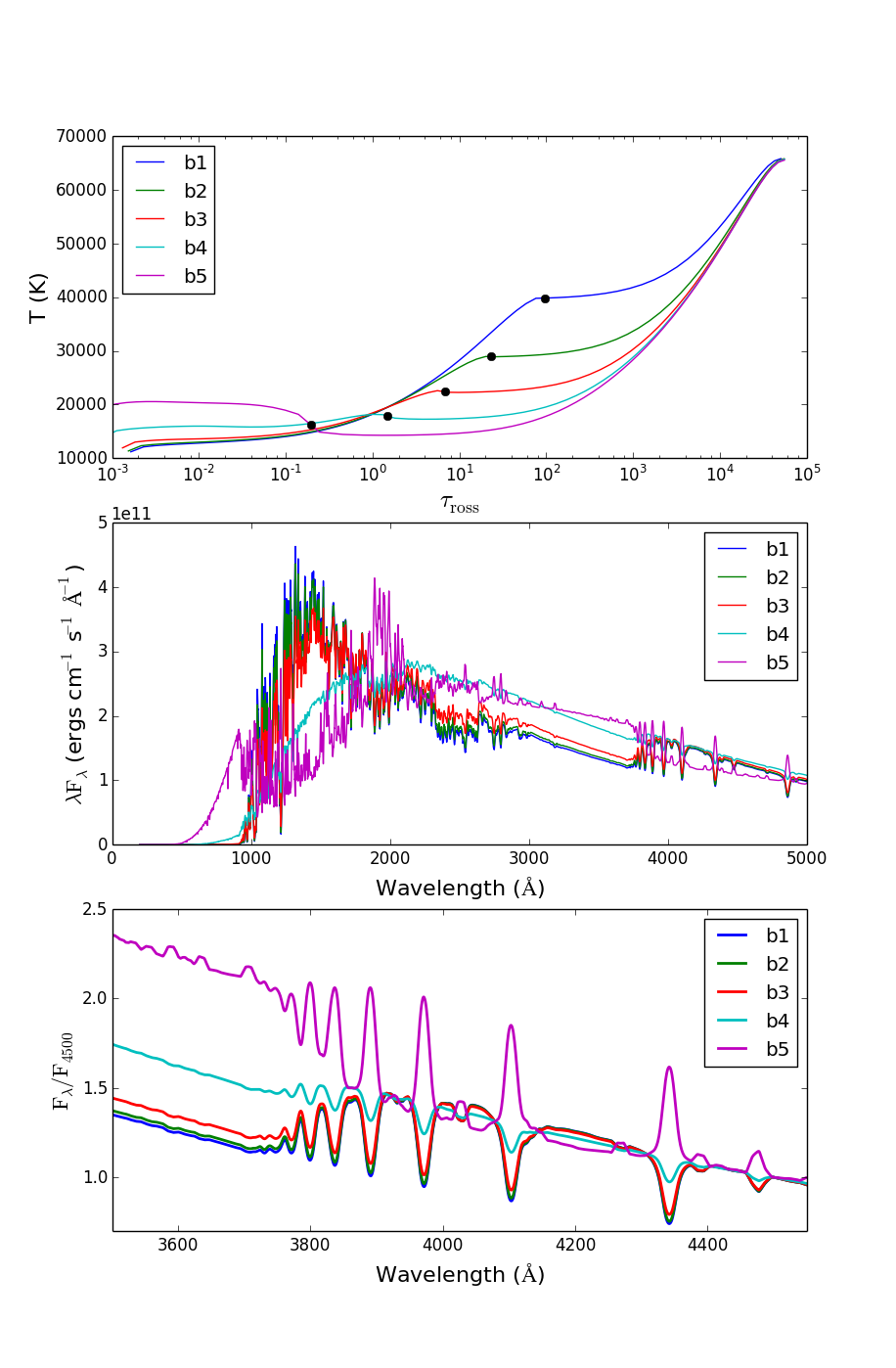}
\caption{Models b1 -- b5: Upper panel: temperature as a function of Rosseland optical depth. Black dots mark the position of the division point $m_d$. Middle panel: emergent flux at the disk surface. Lower panel: zoom on the region of the Balmer discontinuity.  For clarity, we plot here the flux normalized to that at 4500 \AA. The lines representing flux for models b1 and b2 are almost indistinguishable of the plots.
\label{fig:fig2}}
\end{center}
\vspace{-1em}
\end{figure}

\begin{figure}
\begin{center}
\includegraphics[width=3.5in]{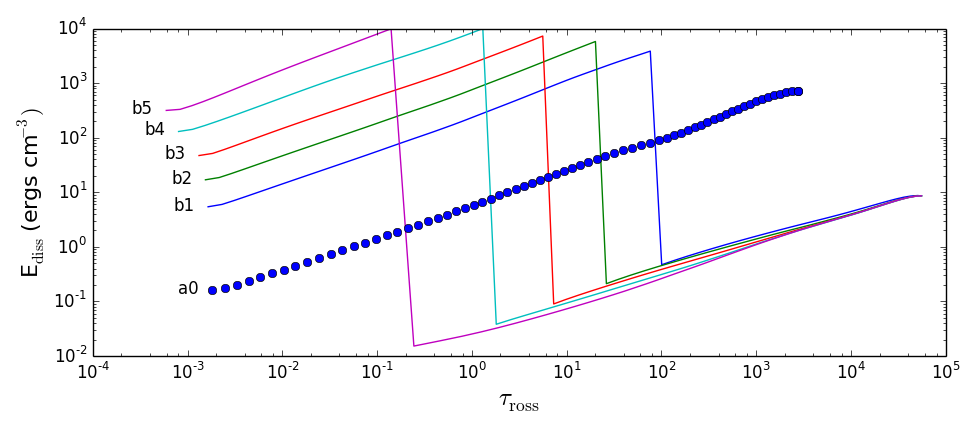}
\caption{Dissipated energy for models b1 -- b5, together with that for the standard model, a0.
\label{fig:fig3}}
\end{center}
\vspace{-1em}
\end{figure}

\begin{figure}
\begin{center}
\includegraphics[width=3.5in]{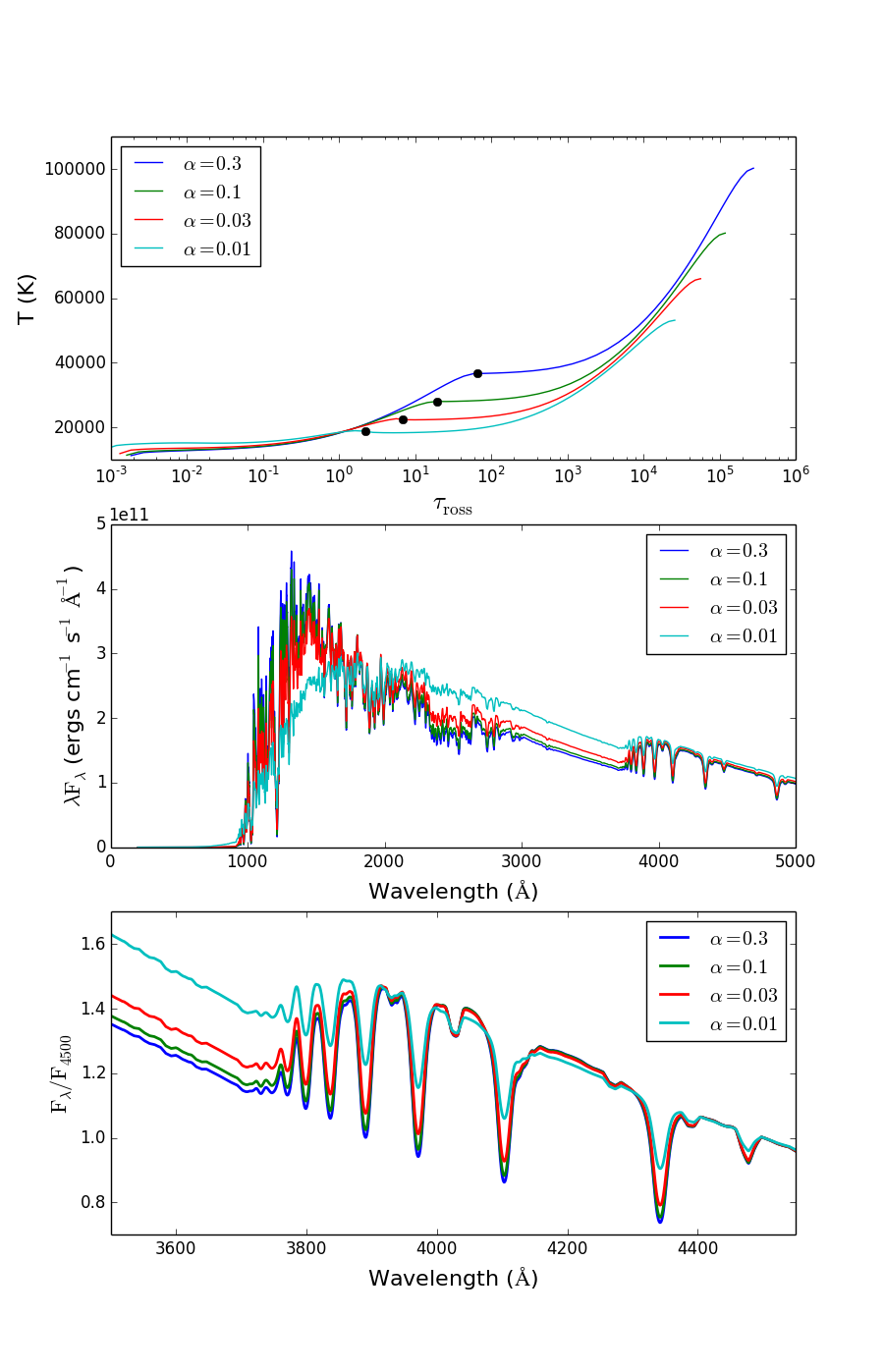}
\caption{A plot analogous to Fig.2, showing a dependence of the temperature structure (upper panel) 
and the emergent flux at the ring surface (middle panel)
on the value of $\alpha$, for constant $f=10^{-2}$, and $m_d/m_0=10^{-3}$. The lower panel is a zoom
on the region of the Balmer discontinuity. Again, for clarity we plot here the flux normalized to that at 4500 \AA.
\label{fig:fig4}}
\end{center}
\vspace{-1em}
\end{figure}

In the first group of models, the a--models, we keep the division point constant at $m_d/m_0 = 10^{-2}$
(which is the adopted standard value in {\sc tlusty}), and vary the fraction $f$ between 0.99 (again, the standard value adopted  in {\sc tlusty}), and $10^{-3}$. The models are displayed in Fig.\,\ref{fig:fig1}.
The temperature in deep layers gradually decreases with decreasing portion of energy dissipated in deep layers, as is to be expected. However, its effect on the emergent flux is very small because the changes in temperature occur at relatively large optical depths, as is clearly seen at the upper panel.

The next group, the b--models, explore the sensitivity of models to the division point. Here, we keep the fraction $f$ constant at $10^{-2}$ and vary the position of the division point. Model b1 is identical to a3, and in the subsequent models the division point $m_d$ is moved upward in order to study its influence on the predicted spectra. Figure \,\ref{fig:fig2}, upper panel,  displays its influence on temperature, essentially flattening the temperature gradient with decreasing $m_d$. 
The model with the lowest division point, $m_d/m_0 = 10^{-4}$ (b5)
exhibits a temperature rise at the surface, a sort of  ``chromosphere". 
Corresponding emergent flux is shown in the middle and the lower panel, which  clearly demonstrate a dramatic effect of the position of the division point on predicted spectra, in particular in the UV region. 
In particular, it is clearly seen that the Balmer jump decreases with decreasing $m_d$, rendering it essentially non-existent for model b4 ($m_d/m_0 =3\times10^{-4}$), and finally turning it into emission for the model b5 ($m_d/m_0 =10^{-4}$), moreover with significant emission cores of high Balmer lines.

Again, the behaviour of the emergent flux is explained by examining the position of the division points, displayed in the upper panel. For models b1 and b2, $m_d$ is located at large optical depths, so the emergent flux is essentially the same for both. Starting with model b3, $m_d$ moves to smaller optical depths, thus having a profound effect on emergent radiation.

To gain more insight, we plot in Fig.\,\ref{fig:fig3}  the dependence of the dissipated energy per unit volume, $E_{\rm diss}$, given by Eq.(\ref{ediss}), on the Rosseland optical depth,  for the b-models, together with that for the standard model (a0). The run
of $E_{\rm diss}$ with depth is clearly reflected in the behavior of the temperature.

\subsection{Sensitivity of models on $\alpha$}

So far, we kept $\alpha$ at the standard value, $\alpha=0.1$. It is well known that the value of
$\alpha$ does not appreciably influence the emergent spectrum for optically thick disks with vertically constant viscosity.
This is because the value of $\alpha$ only influences the total column mass, i.e. the behavior of deep layers, while
the layers around the Rosseland optical depth equal to unity, where the emergent spectrum originates, are influenced only very little. Moreover, the emergent spectrum for such disks is very close to the emergent spectrum for a stellar atmosphere with the same effective temperature, independently of $\alpha$.

However, the situation may be quite different for disk rings for which the viscosity is strongly dependent on depth, such as
those studied in this paper. As shown in Appendix \ref{col_mass}, Eqs. (\ref{scalm0}) and (\ref{betag}), the total column mass scales with $\alpha$ as 
\begin{equation}
m_0 \propto \alpha^{-4/5} f^{1/10}, 
\end{equation}
that is, roughly inversely proportionally to $\alpha$. The temperature
structure is influenced at least between $m_0$ and  $m \approx m_d$. For $m_d/m_0 \ll 1$, which is case studied above, this
leads to significant change in temperature in the layers where the optical depth is close to, or even less than, unity. Consequently, the emergent spectrum may be influenced significantly.

To demonstrate this effect quantitatively, we construct a set of additional models where $f$ and $m_d$ are held fixed, and 
the $\alpha$  parameter is varied between 0.3 and 0.01. The temperature structure is shown in the upper panel of Fig.\,\ref{fig:fig4}, which
clearly illustrates the behavior explained above and in Appendix \ref{col_mass}. The corresponding emergent flux is shown in the middle panel, and a zoom
for the region of Balmer discontinuity in the lower panel. Since the temperature gradient for models with high $\alpha$ is
low, the magnitude of the Balmer jump decreases with increasing $\alpha$.


\begin{figure}
\begin{center}
\includegraphics[width=3.0in]{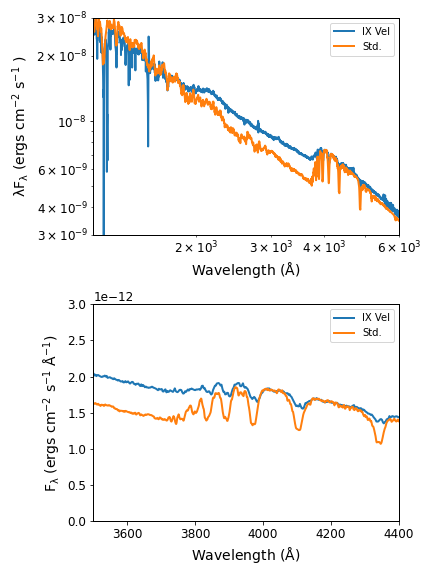}
\caption{Comparison of  synthetic to observed spectra of IX Vel for the standard disk approach were the energy dissipated in the process of accretion is deposited deep within the atmosphere.    \label{fig:ixvel_std}}
\end{center}
\vspace{-1em}

\begin{center}
\includegraphics[width=3.0in]{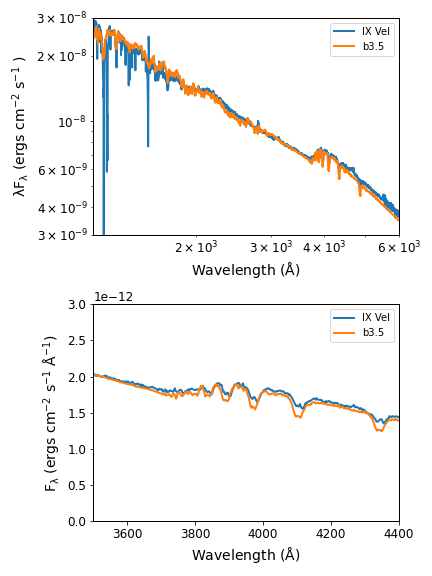}
\caption{Similar to  Fig.\ \ref{fig:ixvel_std}, except the simulated spectra  are for model b3.5 where much of the accretion energy heats the more tenuous surface layers of the disk.    \label{fig:ixvel_new}}
\end{center}
\vspace{-1em}
\end{figure}

\begin{table}
\caption{Parameters for IX Vel}
\begin{center}
\label{tab_ixvel}
\begin{tabular}{lccc}
Parametar & Value & Units & Comment \\
\hline
M$_{wd}$  &  0.8  & \MSOL &  BT90 \\
$ \dot{M}_{disk}$ & 1 $\times$ 10$^{-8}$ & $\MSOL yr^{-1}$ & To match flux \\
R$_{wd} $ & 7.8 $\times$ 10$^{10}$ & cm & M-R relation, P85 \\
R$_{disk}$ & 55  & R$_{wd}$ & BT90 \\
Inclination & 60$\pm$5 & deg & BT90 \\
Distance & 90   & pc &Gaia \\
\hline
\end{tabular}
\\[2.5pt]
{\footnotesize
BT90=\cite{beuermann90}, P85=Paczynski as quoted by \cite{anderson88},  Gaia=\cite{gaia_data}
}
\end{center}
\end{table}

\section{Application to IX Vel}
\label{ixvel}

In the preceding sections, we have shown that the spectra of individual rings of the disk, and in particular  the wavelength range near the Balmer jump are modified by varying the dissipation law that is used. Does this solve the problem when a spectrum for the entire disk is created?  As an initial test, we consider IX Vel, which was  first identified by \cite{garrison84}, which is the brightest cataclysmic variable, and which is, because of its high mass transfer rate, a nova-like variable.  Located at Gaia determined distance of 90.4$\pm$02 pc \citep{gaia_data}, it has a period of 4.7 hr and an inclination of 60$\pm$5$\degr$ \citep{beuermann90}, which implies we have a fairly good view of the accretion disk. 

As a result of its characteristics, such is a very-well observed systems and a number of attempts have been made to model its spectrum, particularly in the UV.   \cite{long94}  made a fairly detailed attempt to fit the HUT (850-1850 \AA) spectrum of IX Vel to steady state disk  models  constructed by summing stellar atmospheres.  The model which fit the  spectral shape  best had a lower $\dot{m}_w$ \EXPU{0.9}{-9}{\MSOL~yr^{-1}} than one needed to match the luminosity, \EXPU{5}{-9}{\MSOL~yr^{-1}} given the value of the distance (95 pc) and inclination (60\degr).   \cite{long94} suggested one possibility was that the temperatures in the inner accretion disk might be less than expected.   \cite{linnell_ixvel} came to very similar conclusions based on FUSE and STIS spectra; they found they could fit the spectra with  $\dot{m}_w$ of \EXPU{5}{-9}{\MSOL~yr^{-1}} it the temperature of the disk within 4 $R_{wd}$ was about 10,000K lower than predicted by the standard steady state disk.

Although nova-like variables are found mostly in the high state, they do vary with time.  Therefore combining spectra obtained in different wavelength bands introduces uncertainties.  To address this problem, we (along with several of our collaborators) used HST to obtain spectra of IX Vel and 4 other nova-like variables covering the wavelength range 1150-10000 $\AA$ using 3 different gratings, as part of program 14637.  Although the spectra were not obtained simultaneously, they were obtained within one HST orbit, and no renormalization of the spectra were required to match the spectra in different wavelength bands.  The shape of the other two non-eclipsing systems are similar to that of IX Vel.  A complete description of these observations along with actual fits to each of the spectra will be given elsewhere.  Our purpose here is simply to show qualitatively that the types of models we are proposing here provide a better match to the spectra than those which ignore energy generation in the atmosphere of the accretion disk.

To create a synthetics spectrum of the accretion disk in IX Vel, we assume that the IX Vel systems has the properties listed in Table \ref{tab_ixvel}.  We divided the disk into 40 approximately logarithmically spaced rings and generated spectra for each ring.
We first construct a ``standard model", equivalent to the above model a0, that is, for $\alpha=0.1$, $m_d=0.01$, and $f=0.99$, which in fact is a model with constant dissipation per volume, and therefore most of the energy is dissipated in the deep, dense, layers, and very little energy in layers with $\tau \approx 1$, so that this model also closely resembles that of a stellar atmosphere.
We then Doppler shifted the spectrum from each ring to allow for its Keplerian rotation and created the total spectrum from an area-weighted sum.  

Guided by the results of the previous section, where we have shown that an acceptable shape of the Balmer discontinuity is somewhere between models b3 and b4, that is, at least for the characteristic ring at $R/R_\ast=15$,
we have constructed a small grid of full disk models and spectra with $\alpha = 0.1$, $f=10^{-2}$, and for several values of $m_d$, namely $3\times 10^{-4}, 5\times 10^{-4},7\times 10^{-4}$, and $1\times 10^{-3}$. It turned out that the best fit
was obtained for $m_d=7\times 10^{-4}$, which we discuss below. For simplicity, we call this model b3.5, i.e. a model between b3 and b4, and for the whole disk.

A comparison of the synthetic spectrum  for a standard disk model at an inclination angle of 60\degr\  to the spectrum of IX Vel is shown in Fig.\  \ref{fig:ixvel_std}.  For the figure, the IX Vel spectrum has been de-reddened assuming E(B-V) = 0.03.  Although this synthetic spectrum resembles that of IX Vel  over limited spectra ranges, it is quite clear that the Balmer jump is too pronounced, the Balmer lines are too deep,  and that the spectral slope does not match the observed spectrum in the region between 1150 - 3100 \AA.  

By contrast, as shown in Fig.\ \ref{fig:ixvel_new}, the integrated synthetic spectrum generated with the above mentioned values of $f=10^{-2}$ and $m_d=7\times 10^{-4}$ and  more closely resembles the actual spectrum of IX Vel than does the synthetic spectrum generated with the standard assumptions.  As expected, the spectrum for the b3.5 model has a less pronounced Balmer jump and weaker Balmer lines, more like that of IX Vel.  Additionally, the slope of the spectrum in the UV is closer to that observed in IX Vel.\footnote{While the statement that the slope is closer to that of IX Vel with model b3.5 than the standard model, there would still be a modest discrepancy between the slope even in model we have chosen had we not dereddened the IX Vel spectrum.}   The relatively narrows features in the observed spectra in the UV are not a concern as they do not arise from the disk. These features arise either from reprocessing of the disk radiation that occurs in a wind emerging from the disk \citep{long94} or interstellar absorption along the line of sight to IX Vel. We conclude, at least in the case of IX Vel, there is no need to adopt modifications to the steady disk model to explain the spectrum.  

We note, in passing, that the results at optical wavelengths, including the Balmer jump are sensitive to the choice of the outer disk radius. There would be obvious differences between the observed spectrum of IX Vel and synthetic spectrum if the outer portion of the disk was truncated at 50 $R_{wd}$.  The strength of the Balmer jump would increase and the model spectra would under-predict the observed spectra at longer wavelengths.

\begin{figure}
\begin{center}
\includegraphics[width=3.3in]{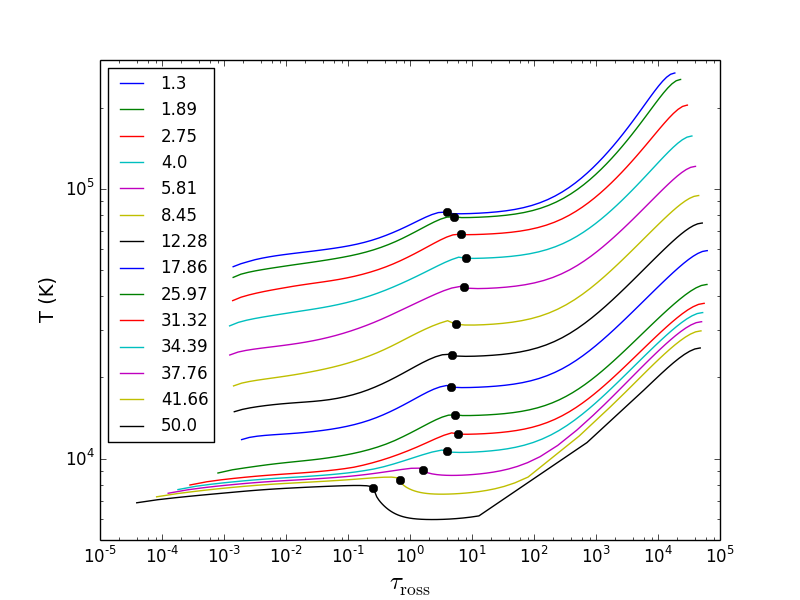}
\caption{Temperature as a function of the Rosseland optical depth for selected rings of the IX Vel disk
model. The black dots indicate the position of the division point $m_d$. 
The curves are labelled by the values of $R/R^\ast$ for the individual rings.
\label{fig:fig7}}
\end{center}
\vspace{-1em}
\end{figure}

For completeness, we present in Fig.\,\ref{fig:fig7}
a plot of $T(\tau_{\rm ross})$ for every fourth ring of the final, best-fit model for IX Vel shown in the previous 
section. For all rings, the division point lies between optical depths 1 and 10; only for the coolest models the
division point lies in the optically thin region.   

\section{Conclusions}
\label{concl}

Despite the fact that nova-like variables and disk-dominated CVs, in the high mass transfer state, are thought to be ``simple'' accretion disk systems, it has been difficult, historically, to create simulated spectra that accurately reproduce the spectra that are observed, particularly over a large spectral range, and especially in the region of the Balmer jump.  Although a variety of modifications to the basic steady-state accretion disk \citep{shakura73}  picture we have of CV disks have been proposed over the years, the most natural has always appeared to us to involve the vertical temperature structure of the disk.  

Here, we have demonstrated a simple parameterized dissipation law, coupled with an improved version of disk/stellar atmospheres structure/radiative transfer code {\sc tlusty}, can be used to produce model spectra that  closely resemble the observed spectra of the brightest nova-like variable, IX Vel.  

The vertical temperature structure in these disk models varies less than for an  analogous stellar atmosphere as well as that of a disk where the energy dissipation occurs mostly near the disk plane.  This results in a less pronounced Balmer jump, weaker Balmer absorption profiles, and a shallower slope in UV, as is observed.  At least in the case of IX Vel, none of the various, in some cases, ad-hoc, modifications to the radial structure of the disk (e.g., changing the inner and/or outer radii) are needed to reproduce the observed spectra.

The parameterized dissipation law that we have used is not based on the micro-physics of accretion disks, but it does suggest that the ultimate physical description of dissipation will result in a significant fraction of the dissipation occurring in the surface layers of the disk.  For the model that that reproduced the spectrum of IX Vel, 99\% of the energy occurs in the surface layers.

The next obvious step is to determine whether this particular prescription for dissipation applies to a larger sample of nova-like and other high mass-transfer rate cataclysmic variables.

\section*{Acknowledgements}

We acknowledge partial support for this project by NASA through grant number HST-GO-14637 from the Space Telescope Science Institute, which is operated by AURA, Inc., under NASA contract NAS 5-26555. KSL acknowledges additional support through grant numbers HST-GO-15984, 16058 and 16066. The IX Vel data was acquired using the NASA/ESA Hubble Space Telescope obtained at the Space Telescope Science Institute.

\section*{Data Availability}

The HST data used here can be retrieved from the HST data archive: \url{http://dx.doi.org/10.17909/t9-81yt-ch19}



\bibliographystyle{mnras}
\bibliography{cv_disk_rev} 




\appendix

\section{Determination of the total column mass of a disk ring}
\label{col_mass}

To determine the column mass at the central plane for a given $\alpha$,  we use Eq. (\ref{mbal}) substituted to (\ref{inttfir}), which yields
\begin{equation}
\label{m0alpha}
m_0 \alpha (\bar P/\bar\rho) = \frac{\dot M}{4\pi}\left(\frac{GM}{R^3}\right)^{1/2}
[1-\left({R_\ast/R}\right)^{1/2}].
\end{equation}
The problem is that $(\bar P/\bar\rho)$ is not known a priori, so that if one intends to use the column mass $m$ as a basic vertical coordinate, one has to devise a procedure to estimate $m_0$ at the outset. 

This problem can be avoided if one uses the Reynolds number instead of $\alpha$ to parameterize the averaged viscosity, as suggested by \cite{lynden74}, and used in constructing vertical structure models -- e.g.,  \cite{kriz86}, \cite{wadehub98}, and others.
However, using the Reynolds number is somewhat dangerous because there is no physically-motivated guidance to
estimate its appropriate values, besides those deduced from hydrodynamical studies of gases under usual Earth-like conditions. For instance, the value suggested by \cite{lynden74}, Re$=5000$, often corresponds to supersonic turbulence in CV disks.\, which is not  realistic.

In the case of dominant radiation pressure, one can easily derive a relation
between $\bar P$ and $\bar\rho$ prior to solving for the detailed structure \cite[e.g.][]{hubeny98}.
In the case where the radiation pressure is not dominant, the (approximate) relation
between the total column mass $m_0$ and the viscosity parameter
$\alpha$ is given by 
\begin{equation}
\label{m0a}
\alpha m_0 \left(\beta_r + \beta_g m_0^{1/4}\right) -\beta_0 =0.
\end{equation}
where
\begin{equation}
\beta_r = (\sigma_R T_{\rm eff}^4 \chi_{\rm e}/c)^2/(3Q),
\end{equation}
\begin{equation}
\beta_g = 0.77 R_g\, \kappa_0^{1/8} (2Q/\pi R_g)^{1/16}\, T_{\rm eff}^{1/2},
\end{equation}
and
\begin{equation}
\label{beta0}
\beta_0 = (\dot M /4\pi) (GM/R^3)^{1/2}\,  [1-\left({R_\ast/R}\right)^{1/2}],
\end{equation}
where $R_g=1.3 \times 10^8$ is the gas constant, 
$\chi_{\rm e}=\sigma_{\rm e}/m_H=0.39$, and $\kappa_0 =6.4\times 10^{24}$ is
the coefficient in the Kramers-type expression for the Rosseland mean opacity,
$\kappa_R \approx \kappa_0 \rho T^{-7/2}$.

Here we provide a detailed derivation of Eqs. (\ref{m0}) -- (\ref{beta0}).
Equation (\ref{m0}) follows from Eqs. (\ref{mbal}) and (\ref{inttfir}), where
one makes an approximation that
\begin{equation}
\label{prhoap}
(\bar P/\bar\rho) \approx (\bar P/\bar\rho)_{\rm rad} + (\bar P/\bar\rho)_{\rm gas},
\end{equation}
where $(\bar P/\bar\rho)_{\rm rad}$ and $(\bar P/\bar\rho)_{\rm gas}$ correspond
to the radiation-pressure dominated and gas-pressure dominated situation, respectively.

The expression for $(\bar P/\bar\rho)_{\rm rad}$ was derived by \cite{hubeny98}.
Briefly, in the case of dominant radiation pressure, and depth-independent viscosity, $\rho$ is
constant with depth, and the hydrostatic equilibrium equation is then $dP/dz=-Q\bar\rho z$, which
has a solution $P(z)=Q\bar\rho(h^2-z^2)/2$. Consequently, $\bar P=Q\bar\rho h^2/3$, and thus
$(\bar P/\bar\rho)_{\rm rad}= Q h^2/3$. The disk height $h$ is given by [see Eq. (53) of \cite{hubeny98}].
\begin{equation}
h = \frac{\sigma_R T_{\rm eff}^4}{Q} \frac{\kappa_H}{c},
\end{equation}
where $\kappa_H$ is the flux-mean opacity. It can be approximated by the Rosseland mean opacity, which in
turn is further approximated by the electron scattering  opacity, $\sigma_{\rm e}/m_H$, that one finally obtains
\begin{equation}
\label{prhorad}
(\bar P/\bar\rho)_{\rm rad}= (\sigma_R T_{\rm eff}^4 \chi_{\rm e}/c)^2/(3Q).
\end{equation}

\medskip

In the case the gas pressure is dominant, we have
\begin{equation}
P(z)/\rho(z) = R_g T(z),
\end{equation}
so that the vertically-averaged $P/\rho$ is directly proportional to the vertically averaged temperature.
An analytic expression for $T$ as a function of a mean opacity (approximated again by the Rosseland mean opacity)
was derived by \cite{hubeny90}. We use here a simplified version of the general expression, valid for large total
optical thickness of the disk, namely
\begin{equation}
\label{temp}
T(\tau)^4 = \frac{3}{4} T_{\rm eff}^4 \left[\tau \left(1-\frac{\tau}{2\tau_c}\right) +\frac{1}{\sqrt{3}} +
\frac{1}{3\epsilon \tau_c}\right] 
\end{equation}
where $\tau$ is the Rosseland optical depth, $\tau_c$ is the optical depth at the central plane, and
$\epsilon = \kappa_B/\kappa_R$ is the ratio of the Planck-mean to the Rosseland mean opacity.
For large optical depth, $\tau_c\gg 1$, the two last terms in square bracket are negligible. 
The average $P/\rho$ can then be written as
\begin{equation}
(\bar P/\bar\rho)_{\rm gas} = (R_g/h) \int_0^h\!\! T(z)dz = (R_g/\tau_c) \int_0^{\tau_c}\!\! T(\tau) d\tau.
\end{equation}
and therefore
\begin{eqnarray}
\label{prhog}
(\bar P/\bar\rho)_{\rm gas} &=& (R_g/\tau_c) (3/4)^{1/4} T_{\rm eff} \int_0^{\tau_c} [\tau(1-\tau/2\tau_c)]^{1/4} d\tau \nonumber \\
&=& R_g (3/4)^{1/4} T_{\rm eff}\, I\, \tau_c^{1/4},
\end{eqnarray}
where $I = \int_0^1 x^{1/4}(1-x/2)^{1/4} dx \approx 0.735$.
It now remains to estimate the optical depth at the central plane, $\tau_c$. It is given by
\begin{equation}
\tau_c = \int_0^{m_0} \kappa_R (m) dm \approx m_0 \kappa_R(m_0),
\end{equation}
where we have approximated a generally depth-dependent $\kappa_R$ by its value at the central plane.
Using the Kramers-type expression, $\kappa_R = \kappa_0 \rho T^{-7/2}$, we obtain
\begin{equation}
\tau_c = m_0 \kappa_0 \rho_c T_c^{-7/2},
\end{equation}
where $\rho_c$ and $T_c$ are the density and temperature at the central plane. From the isothermal hydrostatic
equilibrium equation one has $\rho_c = m_0(2Q/\pi c_s^2)^{1/2}$, where $c_s$ is the isothermal sound speed, so that
$\rho_c = m_0 (2Q/\pi R_g)^{1/2} T_c^{-1/2}$, and therefore
\begin{equation}
\tau_c = m_0^2 \kappa_0 \left(\frac{2Q}{\pi R_g}\right)^{\!\!1/2} T_c^{-4}.
\end{equation}
The central temperature is given, using Eq. (\ref{temp})
\begin{equation}
T_c^4 = (3/8) T_{\rm eff}^4 \, \tau_c,
\end{equation}
so that one obtains the following expression for $\tau_c$,
\begin{equation}
\tau_c = m_0^2 \kappa_0 \left(\frac{2Q}{\pi R_g}\right)^{\!\!1/2} (3/8)^{-1} \, T_{\rm eff}^{-4} \tau_c^{-1},
\end{equation}
from which one obtains a desired expression for $\tau_c$,
\begin{equation}
\label{tauc}
\tau_c = m_0 \kappa_0^{1/2} \left(\frac{2Q}{\pi R_g}\right)^{\!\!1/4} (3/8)^{-1/2}  T_{\rm eff}^{-2}.
\end{equation}
Substituting this to Eq. (\ref{prhog}), one obtains
\begin{equation}
\label{prhogas}
(\bar P/\bar\rho)_{\rm gas} = m_0^{1/4} C R_g \kappa_0^{1/8} \left(\frac{2Q}{\pi R_g}\right)^{\!\!1/16}T_{\rm eff}^{1/2},
\end{equation}
where the constant $C=(3/4)^{1/4} (3/8)^{-1/8} I \approx 0.77$.

Substituting Eqs. (\ref{prhogas}), (\ref{prhorad}), and (\ref{prhoap}) into (\ref{inttfir}), one obtains the
equation for $m_0$ that contains only input model parameters, Eq. (\ref{m0}).
This non-linear equation for $m_0$ is solved by the Newton-Raphson method. 

If the gas pressure is dominant, the solution of Eq. (\ref{m0}) is simply
\begin{equation}
\label{scalm0}
m_0 = \left(\frac{\beta_0}{\alpha\beta_g}\right)^{\!4/5}.
\end{equation}

The above derived approach, suggested already by \cite{hubeny17}, is valid only for a depth-independent viscosity, 
$w(m) = \bar w$. Since we have explored here an influence of a  step-wise viscosity defined through the parameters $f$ and $m_d$, we need to provide a generalization of this approach. Since the formalism for general values of $f$ and $m_d$  is unnecessarily complicated, we limit ourselves to the case $f \ll 1$ and $m_d/m_0 \ll 1$, which are indeed the cases studied in 
this paper.

The most significant change is in Eq. (\ref{temp}). A more general expression for the temperature was derived by  \cite{hubeny90},
\begin{equation}
\label{tempf}
T(m)^4 = \frac{3}{4} T_{\rm eff}^4\left[\tau_H(m)-\tau_\theta(m) + \frac{1}{\sqrt{3}}\right],
\end{equation}
where
\begin{equation}
\label{tauhm}
\tau_H(m)=\int_0^m\kappa_H(m^\prime) dm^\prime, \quad 
\end{equation}
and
\begin{equation}
\tau_\theta(m)=\int_0^m\kappa_H(m^\prime) \theta(m^\prime) dm^\prime,
\end{equation}
where $\kappa_H$ is the flux-mean opacity, approximated by the Rosseland mean opacity, and
$\theta$ is defined by Eq. (\ref{theta}). For the viscosity given through Eq. (\ref{visc2}), $\theta$ is given,
for $m>m_d$, by
\begin{equation}
\theta(m) = (1-f)+f\frac{m-m_d}{m_0-m_d} \approx f \frac{m}{m_0}.
\end{equation}
Consequently, Eq. (\ref{tempf}) can be approximately written, in analogy to Eq. (\ref{temp}), as
\begin{equation}
T(\tau)^4 = \frac{3}{4} T_{\rm eff}^4\left[\tau\left(1-\frac{\tau}{2\tau_c}\right) f + \frac{1}{\sqrt{3}}\right].
\end{equation}
Using this expression for $T$, one modifies Eq. (\ref{prhog}) to
\begin{equation}
(\bar P/\bar\rho)_{\rm gas} \approx R_g (3/4)^{1/4} T_{\rm eff}  I \tau_c^{1/4} f^{1/4},
\end {equation}
and Eq. (\ref{tauc}) to
\begin{equation}
\tau_c = m_0 \kappa_0^{1/2} \left(\frac{2Q}{\pi R_g}\right)^{\!\!1/4} (3/8)^{-1/2}  T_{\rm eff}^{-2} f^{-1/2}.
\end{equation}
Finally, Eq. (\ref{m0}) remains valid, with the coefficient $\beta_g$ now given by
\begin{equation}
\label{betag}
\beta_g = 0.77 R_g\, \kappa_0^{1/8} (2Q/\pi R_g)^{1/16}\, T_{\rm eff}^{1/2} f^{-1/8},
\end{equation}

\medskip

Since the above described procedure is based on a number of approximations, we improve the model by the following procedure: We construct a vertical structure model for a given ring using the estimate of $m_0$ using Eq. (\ref{m0}). Once the model is computed, we also calculate an ``equivalent $\alpha$" that corresponds to $m_0$ and actual $\bar P$ and $\bar\rho$ of the computed model, using Eq. (\ref{m0alpha}), which is now viewed as an equation for $\alpha$. If the original $\alpha$ and newly computed equivalent $\alpha$ agree, the model is correct. If not, one recalculates the model with a different input $\alpha$, which is the original $\alpha$, scaled by the ratio of the equivalent to the original $\alpha$. If needed, one can iterate this procedure, but in our experience, just one recalculation of the modes is completely sufficient.


\section{Treatment of hydrostatic equilibrium within the Rybicki scheme}
\label{rybicki} 

As mentioned above, the Rybicki scheme is described in detail elsewhere. Here we only mention
that the scheme is essentially a complete linearization where a global structure is reversed from
the original grand Jacobi matrix. from being essentially tridiagonal (corresponding to discretizing
second-order differential equations), with the individual blocks being diagonal matrices with several additional
rows and columns. Here, only the energy balance equation is being solved simultaneously with the
radiative transfer equation, so that there is just one additional row and column in otherwise diagonal matrices.
When the structure is reversed, as is done in the Rybicki scheme, the outer, global, matrix is block-diagonal, with
just one additional row and column of blocks, where the blocks now being tri-diagonal, and the 
whole numerical performance consist of numerous inversions of a tri-diagonal matrices, which is very cheap and fast Therefore, one can consider a large number of discretized frequency points, and therefore represent 
the radiation field more accurately.
Moreover, since the method is essentially a complete linearization, the convergence is usually very fast.

However, only the radiative transfer equation and the energy balance equation are solved simultaneously.
In other words, only new temperature, and new radiation intensities follow directly from the linearization process. This means that other structural equations, in particular the hydrostatic equilibrium equation, must be solved separately to determine other structural parameters, such as the pressure and the density, corresponding to the new temperature.

This is not a problem in the case of stellar atmospheres, and in particular for cool stars where the radiation pressure is negligible. Then $P=mg$ is invariant, so that the total particle number density corresponding to the new temperature is given by $N=P/kT$. One then determines the electron density and all the atomic level populations by a standard procedure of solving the set of Saha-Boltzmann equations. For details, refer to \cite{hubmih14}. If the radiation pressure is not negligible, then the new radiation pressure is estimated as
$$
P_{\rm rad}^{\rm new} = P_{\rm rad}^{\rm old}\, (T_{\rm new}/T_{\rm old})^4,
$$
which follows from the assumption that departures of radiation pressure from its equilibrium value remain unchanged between iterations.

The situation is more complicated in the case of disks, because the gravity acceleration is proportional 
to the vertical distance from the central plane, $z$ -- see Eqs (\ref{hedisk}) and
(\ref{gdef})
\begin{equation}
\label{qdef}
g(z) = Q z, \quad Q\equiv GM_\ast/R^3,
\end{equation}

In {\sc tlusty}, the column mass is a basic geometrical coordinate, while $z$ is taken as one of the
structural parameters to be determined by the model. It is given by the so-called $z$-$m$ relation.
\begin{equation}
\label{zm}
dm = -\rho dz,
\end{equation}
where $\rho$ is the mass density. The minus sign follows from the convention that $z$ is taken as 0 at 
the central plane and increases upward, while $m$ increases from the top downward. 

To update the gas pressure, density, and the vertical distance for a new $T$,
we use an
approach suggested by \cite{hubeny90}. Here the dependence of $g$ on $z$ is accounted for by
differentiating Eq. (\ref{hedisk}) once more over $m$, and using Eq. (\ref{zm}) to obtain a second-order equation
for $P_{\rm gas}$, namely
\begin{equation}
\label{p2}
\frac{d^2\! P_{\rm gas}}{dm^2} = - \frac{Q}{\rho} - \frac{d^2\! P_{\rm rad}}{dm^2}.
\end{equation}
We then express $\rho$ using a relation 
\begin{equation}
P_{\rm gas} = c_s(T)^2 \rho = \gamma T  \rho.
\end{equation}
Here, the first equality expresses the mass density through the isothermal sound speed, while the
second equality takes the square of the sound speed proportional to the temperature. 
Notice also that if the gas pressure is dominant, $\gamma \rightarrow R_g$, the gas constant.
Equation (\ref{p2})
is then written as (writing $P_{\rm gas}$ simply as $P$)
\begin{equation}
\label{prad22}
P \frac{d^2 P}{dm^2} = -\gamma T Q - \frac{d^2 P_{\rm rad}}{dm^2}.
\end{equation}
where the quantities $\gamma$ and $d^2 P_{\rm rad}/dm^2$ are taken from the previous iteration step.
An advantage of  using
$d^2P_{\rm rad}/dm^2$ from the previous iteration step
follows from the observation that this quantity does not change very much from iteration to iteration. It is seen
by expressing
\begin{equation}
\frac{d^2 P_{\rm rad}}{dm^2} = \frac{d g_{\rm rad}}{dm} = 
\frac{4\pi}{c} \frac{d}{dm}\left( \int_0^\infty\!\! \chi_\nu H_\nu d\nu \right)
\approx \frac{4\pi}{c} \frac{d(\chi_{\rm Ross} H)}{dm} 
\end{equation}
where $\chi_{\rm Ross}$ is the Rosseland mean opacity, and $H$ the total flux. $H$ does not change from iteration
to iteration, and $\chi_{\rm Ross}$ changes only a little, so our approach of keeping the second derivative of
the radiation pressure from the previous iteration step is quite reasonable.

The lower boundary condition follows from expressing the pressure at depth $D-1$ ($D$ being the 
index of the last depth point corresponding to the central plane) through a Taylor expansion (writing $m_D$
as $M$)
\begin{eqnarray}
\label{hubc}
P(m_{D-1}) = P(M) + (m_{D-1} - M) P^\prime (M) + \nonumber \\å
(1/2) (m_{D-1} - M)^2 P^{\prime\prime}(M),
\end{eqnarray}
where $P^\prime \equiv dP/dm$. Here, $P^\prime(M)=0$ because of the symmetry of the disk around the midplane.
The lower boundary condition follows from substituting  the second derivative of $P$ 
from Eq. (\ref{p2}) to Eq. (\ref{hubc}). 

The upper boundary condition is more complicated. It was  derived by \cite{hubeny90}; here we present only the
final result:
\begin{equation}
\rho_1 = \frac{m_1}{H_g f_1}.
\end{equation}
where
\begin{equation}
f_1 = f\left(\frac{z_1-H_r}{H_g}\right)\!, \quad {\rm with}\quad\quad f(x) \equiv \frac{\sqrt{\pi}}{2} \exp(x^2)\, {\rm erfc}(x),
\end{equation}
and
\begin{equation}
H_g = \left(\frac{2\gamma_1 T_1}{Q}\right)^{1/2}\!\!\!, \quad {\rm and} \quad H_r = \frac{g_{{\rm rad}, 1}}{Q}.
\end{equation}
Upon discretizing, one obtains a set of non-linear algebraic equations for $P_d$, viz,
\begin{eqnarray} 
\label{trip}
P_1 - \frac{m_1}{f_1} \left(\frac{\gamma T_1 Q}{2}\right)^{1/2} &=& 0, \nonumber \\
a_d P_{d-1}P_d  + c_d P_{d+1}P_d  - b_d P_d^2  + q_d P_d  + \gamma_d T_d Q &=& 0 , \nonumber \\ d=2,\ldots,D-1, \\
P_{D-1} P_D \frac{2}{\Delta m_{D-1/2}^2} - P_D^2 \frac{2}{\Delta m_{D-1/2}^2} + \gamma_D T_D Q &=& 0 \nonumber,
\end{eqnarray}
where $q_d \equiv (d^2 P_{\rm rad}/dm^2)_d$, and
\begin{eqnarray}
a_d &=& \frac{1}{\Delta m_{d-1/2}\Delta m_d}, \\
c_d &=& \frac{1}{\Delta m_{d+1/2}\Delta m_d},\\
b_d &=& a_d + c_d,
\end{eqnarray}
for $d=2,\ldots, D-1$, with
\begin{equation}
\Delta m_{d\pm 1/2} = |m_d - m_{d\pm 1}|, \quad \Delta m_d = (m_{d+1/2} - m_{d-1/2})/2.
\end{equation}
The set of non-linear equations (\ref{trip}) is solved by the standard Newton-Raphson method. Once the new gas pressure is determined, the procedure is the same as in the case of stellar atmospheres, i.e. determining the new $N$ and then from the Saha-Boltzmann equations new electron density and atomic level populations.


\bsp	
\label{lastpage}
\end{document}